\documentclass[a4paper,fleqn]{cas-dc}
\usepackage[numbers,sort&compress]{natbib} 
\usepackage{bm}
\usepackage{epstopdf}
\usepackage[sf]{subfigure}
\usepackage{hyperref}
\hypersetup{
	colorlinks=true,
	linkcolor=cyan,
	filecolor=magenta,      
	urlcolor=blue,
}
\usepackage[most]{tcolorbox}

\begin{document}
\shorttitle{A Dimensionally Consistent Size–Strain Plot Equation}
\shortauthors{Anand Pal, \today}

\title [mode = title]{A Dimensionally Consistent Size–Strain Plot Method for Crystallite Size and Microstrain Estimation}

\author[1]{Anand Pal}[orcid=0000-0003-1602-507X]
\cormark[1]
\ead{sandhu.anand@hotmail.com}
\credit{Conceptualization, Methodology, Validation, Formal Analysis, Investigation, Writing – Original Draft, Writing – Review \& Editing}
\address[1]{Department of Physics, Manipal Institute of Technology Bengaluru, Manipal Academy of Higher Education, Manipal-576104, India}

\begin{abstract}
	X-ray diffraction (XRD) peak broadening analysis remains a cornerstone for quantifying crystallite size and lattice microstrain in materials. Among various approaches, the Size–Strain Plot (SSP) method is widely employed for its conceptual simplicity and ease of use. However, this study reveals that the equation most commonly applied in SSP analysis is dimensionally inconsistent—a critical flaw that has gone largely unnoticed and replicated across decades of materials research. This pervasive error raises concerns about the validity of a significant body of published microstructural data. By tracing the historical origin of the misformulated equation, we demonstrate how a seemingly minor oversight evolved into a widely accepted standard practice within the field. We then present a dimensionally consistent formulation that restores physical meaning and analytical reliability to the SSP method. The corrected framework re-establishes the SSP approach as a robust and physically valid tool for XRD-based microstructural characterization.
\end{abstract}


\maketitle

\section{Introduction}
A century after the Scherrer equation first linked X-ray diffraction (XRD) peak broadening to crystallite size, XRD has evolved into one of the most versatile and indispensable tools in materials research \cite{Scherrer1918}. Beyond phase identification, XRD enables detailed insights into the structural and microstructural parameters of materials, such as lattice parameters, preferred orientation, residual stress, defect density, crystallite size, and lattice microstrain \cite{Cullity14,Yung_Rietveld}. Due to these reasons, the twentieth century witnessed extensive efforts by materials scientists to understand X-ray diffraction peak broadening and its correlation with crystallite size and lattice microstrain. Since the formulation of the Scherrer equation in 1918, several refinements have been proposed to improve the quantitative interpretation of diffraction profiles. The Warren–Averbach (WA) method, based on Fourier analysis, provides the most rigorous framework for separating size and strain effects. However, its mathematical complexity and stringent data-quality requirements have restricted its widespread adoption \cite{Warren1950}

Williamson and Hall introduced a graphical approach based on the integral breadth of diffraction peaks to simplify the analysis to estimate size and strain \cite{Hall1949,Williamson1953}. Nevertheless, the Williamson–Hall (W–H) method often yields inaccurate size–strain separation because it assumes inappropriate peak profile shapes\cite{Williamson1953}. To overcome the limitations of both the mathematically intensive Fourier approach and the oversimplified W–H plot, the Size–Strain Plot (SSP) method was developed as a graphical approach that treats the diffraction-line profile as a convolution of Gaussian and Lorentzian components, providing a more physically realistic representation while retaining computational simplicity \cite{Halder1965,Langford1992}. This balance between physical accuracy and simplicity has led to the widespread adoption of the SSP approach in the materials science community.

However, despite its widespread adoption and perceived reliability, the most frequently employed formulation of the SSP equation is dimensionally inconsistent, resulting in physically nonsensical outcomes \cite{Zak2011,Bindu2014,Prabhu2014,RajeshKumar2017,Hassanzadeh-Tabrizi2023}. What likely began as a minor typographical or algebraic oversight has propagated through the literature for over a decade, reinforced by repetition in highly cited publications. This flawed equation has been applied in thousands of studies, particularly in the field of nanoscience, leading to quantitatively inaccurate and physically meaningless results. Surprisingly, this error has gone largely unnoticed for nearly two decades, raising concerns about the validity of existing microstructural data and highlighting limitations in the critical self-correction mechanisms within the scientific community.

This study demonstrates how a seemingly minor oversight, when repeatedly built upon, can distort an entire analytical framework. We first examine the problem with the widely used formulation of the Size–Strain Plot (SSP) equation and then present a brief historical overview of size and strain analysis to place the issue in context. Subsequently, we trace the origin of the misformulated SSP equation, showing how an early dimensional inconsistency propagated through the literature over nearly two decades. Based on this historical and analytical review, we propose a dimensionally consistent form of the SSP equation and critically correct a long-standing methodological error in the materials science community. A preliminary account of this issue was earlier presented in a short comment published in Solid State Sciences, where the dimensional inconsistency was briefly noted \cite{Pal23}. 

The present work provides a complete account of the problem, its historical development, and a corrected formulation. By addressing this methodological inconsistency, we aim to restore rigor to XRD-based microstructural analysis and highlight the importance of continuous scrutiny—even of widely accepted and routinely applied analytical tools.

\section{Historical Background of X-ray Line-Profile Analysis}

A brief historical overview of X-ray diffraction line-profile analysis is essential to place the present work in context before discussing the origin of the dimensional inconsistency and proposing the corrected formulation. The earliest quantitative description, proposed by Scherrer in 1918 \cite{Scherrer1918}, established that finite crystallite dimensions lead to diffraction peak broadening inversely proportional to crystallite size:
\begin{equation}
	\beta_{hkl} = \frac{K\,\lambda}{D \cos\theta},
	\label{eq:Scherrer}
\end{equation}
where $\beta_{hkl}$ is the peak breadth, $\lambda$ is the X-ray wavelength, $D$ is the crystallite size, $\theta$ is the Bragg angle, and $K$ is a dimensionless shape factor. Although Scherrer’s treatment assumed idealized crystallite shapes, it provided the conceptual foundation for subsequent developments in line-profile analysis.

Early experimental observations revealed that peak broadening could not be attributed solely to finite domain size. Lattice distortions (microstrain) arising from cold working also contribute significantly to diffraction peak broadening, with distinct angular dependencies for size and strain effects \cite{Dehlinger1927}. Dehlinger and Kochendörfer (1939) showed that the size broadening varies as $1/cos\theta$ and strain broadening as $tan\theta$ \cite{Dehlinger1939}. Their results demonstrated that coherently diffracting domain size remained constant under deformation, while strain increased systematically.

A major methodological advance came with the Fourier-based approach of Warren and Averbach (1950), which enabled the extraction of size and strain information directly from the Fourier coefficients of entire diffraction profiles without assuming a particular peak shape \cite{Warren1950}. This method provided detailed insight into the non-uniform nature of strain in cold-worked metals. However, its mathematical complexity, sensitivity to background subtraction, and stringent data-quality requirements limited its widespread application.

In parallel, Williamson and Hall introduced a more straightforward graphical approach, emphasizing the limitations of single-criterion methods, such as simple $\tan \theta$  and $\sec \theta$ plots, in reliably separating size and strain contributions. They noted that such approaches can lead to unphysical results, including negative intercepts for very small crystallites \cite{Williamson1953}. As an alternative, Williamson and Hall proposed that the total broadening arising from both size and strain effects can be expressed as
\begin{equation}
	\beta_{hkl} \cos \theta = \frac{K\lambda}{D} + 2\sigma \sin \theta,
	\label{eq:WHequ}
\end{equation}
where the intercept and slope of the linear plot provide estimates of the crystallite size and microstrain, respectively.

However, Williamson and Hall also emphasized that real diffraction profiles seldom conform strictly to either Lorentzian (Cauchy) or Gaussian shapes; rather, they are typically a convolution of both. Consequently, conventional analyses that assume purely additive breadths (Cauchy–Cauchy) or additive squares (Gaussian–Gaussian) fail to capture the true nature of the diffraction line profiles. This distinction is significant because size broadening is predominantly associated with Lorentzian profiles, whereas strain broadening is more accurately described by Gaussian or related distributions \cite{Langford1971,Langford1980}.
The observed diffraction profile $I_{\mathrm{obs}}(x)$ is well described as the convolution of a Lorentzian (Cauchy) function ($I_C(x) $), representing size broadening, and a Gaussian function ($I_G(x) $), representing strain broadening \cite{Langford1978}:
\begin{equation}
	I_{\mathrm{obs}}(x) = \int_{-\infty}^{\infty} I_C(x-u)\,I_G(u)\,du.
	\label{eq:convoLG}
\end{equation}
Schoening derived an \emph{exact} expression for the integral breadth of this Voigt profile \cite{Schoening1965}:
\begin{equation}
	\beta = \beta_G\,\frac{\exp\!\left[-\left(\frac{\beta_C}{\sqrt{\pi}\,\beta_G}\right)^2\right]}{1-\operatorname{erf}\!\left(\frac{\beta_C}{\sqrt{\pi}\,\beta_G}\right)},
	\label{eq:voigt}
\end{equation}
where $\beta_C$ and $\beta_G$ are the Lorentzian and Gaussian components, respectively, and $\operatorname{erf}$ is the error function. Although theoretically rigorous, the complexity of this expression and its reliance on specialized computational functions limited its applicability for routine line-profile analysis at the time. To overcome this challenge of evaluating the exact Voigt convolution, Halder and Wagner proposed an algebraic approximation linking the total peak breadth $\beta$ to its Lorentzian and Gaussian components, $\beta_C$ and $\beta_G$ \cite{Halder1965,Halder1966}:
\begin{equation}
	\beta^{2} = \beta_{C}\,\beta + \beta_{G}^{2}.
	\label{eq:viogtapprox}
\end{equation}
This approximation remained accurate to within approximately 5\% \cite{Langford1978} and formed the basis of the Size--Strain Plot (SSP) method.  

However, Eq.~(\ref{eq:viogtapprox}) did not constitute a peak-shape function; it did not describe the diffraction profile and could only be applied after the peak had been fitted with an analytical function that provided $\beta$, $\beta_C$, and $\beta_G$.  
For this reason, analytical approximations to the Voigt profile were needed that offered both accuracy and computational efficiency. A variety of analytical profile shape functions have been utilized for modeling X-ray diffraction peaks, including Gaussian (G), Lorentzian (L), modified Lorentzian (ML), intermediate Lorentzian (IL), Cosine-Lorentz, pseudo-Voigt (pV), Pearson VII, polynomial ratio, and Student's type functions. Among these, the Gaussian function consistently provides the least satisfactory fit, whereas the pseudo-Voigt and Pearson VII profiles are the most effective in closely approximating observed X-ray diffraction peak shapes \cite{Young1982}.

The pseudo-Voigt (pV) function was proposed as a practical approximation of the Voigt profile \cite{Whiting1968}. It represented the peak as a linear combination of Lorentzian and Gaussian components with a common full width at half maximum (FWHM) \cite{Kielkopf1973, Wertheim1974}:
\begin{equation}
	\mathrm{pV}(x) = I_{0}\left[\eta\,L(x,\beta) + (1-\eta)\,G(x,\beta)\right],
\end{equation}
where $\eta$ was the Lorentzian fraction ($0 \leq \eta \leq 1$) and $\beta$ was the shared FWHM.  For $\eta = 0$, the function reduced to a pure Gaussian, and for $\eta = 1$, it reduced to a pure Lorentzian. The pV function approximated the Voigt profile to within about 1\% and offered a substantial reduction in computational cost \cite{Wertheim1974}.

Hall \textit{et al.}\ \cite{Hall1977} demonstrated that the Pearson~VII function provides an analytical peak profile capable of fitting symmetric X-ray diffraction peaks using a single shape parameter, $m$:
\begin{equation}
	y(x) = y_{0}\left[1 + \frac{(x-\bar{x})^{2}}{m a^{2}}\right]^{-m}.
\end{equation}
The parameter $m$ controlled the peak form, varying continuously from Lorentzian ($m=1$) to Gaussian ($m\rightarrow\infty$). Although Pearson~VII described a wide range of symmetric profiles, the parameter $m$ lacked direct physical correspondence to Gaussian and Lorentzian broadening contributions, and the exponent made the function more computationally demanding than pseudo-Voigt.

Thompson, Cox, and Hastings (TCH) refined the pseudo-Voigt approximation by expressing it in terms of the individual Gaussian and Lorentzian widths, $\beta_G$ and $\beta_C$, rather than the mixing parameter $\eta$~\cite{Thompson1987}. This formulation improved physical interpretability because $\beta_G$ and $\beta_C$ could be associated directly with strain and size broadening. The pseudo-Voigt (TCH) function subsequently became the standard analytical peak-shape model in Rietveld refinement programs \cite{Rodrguez-Carvajal1993, TOPAS5Manual}.

To compare the behaviour of these functions, we fitted a representative diffraction peak using Voigt, pseudo-Voigt, pseudo-Voigt (TCH), and Pearson~VII models (Figure~\ref{fig:peakfits}). 
All four functions accurately reproduced the central region of the peak. However, the Pearson~VII function showed noticeable deviations in the tail region, as highlighted in the inset. Residual analysis confirmed that the Voigt, pseudo-Voigt, and pseudo-Voigt (TCH) functions provided the most accurate description of the experimental data.  
Among these, the TCH formulation offered the practical advantage of yielding $\beta_C$ and $\beta_G$ directly, facilitating subsequent size–strain analysis.
\begin{figure}
	\centering
	\includegraphics[width=0.9\linewidth]{Figure1.pdf}
	\caption{Peak-shape fitting of a representative XRD reflection using four analytical functions: Voigt, pseudo-Voigt, pseudo-Voigt (Thompson--Cox--Hastings, TCH), and Pearson~VII. The upper panel shows the experimental data (black dots) and the fitted profiles (solid lines), normalised to the same peak height for clarity. 
		The lower panel presents the corresponding residuals ($y_{\rm obs} -y_{cal}$), 
		highlighting that the Voigt, pseudo-Voigt, and pseudo-Voigt(TCH) functions provide the closest agreement with the experimental peak shape,   whereas the Pearson~VII profile exhibits noticeable deviations in the tail region.}
	\label{fig:peakfits}
\end{figure}

\section{Re-examining the Widely Used SSP Equation}
The version of the Size–Strain Plot (SSP) equation most widely used in the literature has appeared independently in several influential publications over the past two decades and has since become the de facto formulation for microstructural analysis. Notable early examples include the works of Zak \emph{et al.} (2011, cited over 2784 times) \cite{Zak2011, GoogleScholar}, Bindu \emph{et al.} (2014, cited over 1724 times) \cite{Bindu2014, GoogleScholar}, and Prabhu \emph{et al.} (2014, cited over 774 times) \cite{Prabhu2014, GoogleScholar}. These studies and many others have contributed to a rapidly expanding body of literature built on the same underlying formulation.

This collective usage has resulted in what may be described as a citation pyramid, where repeated citation and reuse of the same equation have substantially amplified its research. For instance, Debojyoti Nath \emph{et al.} (2020) \cite{Nath2020} and a recent review by S.A. Hassanzadeh-Tabrizi (2023) \cite{Hassanzadeh-Tabrizi2023} both cite the work of Zak \emph{et al.} \cite{Zak2011}; these papers have accumulated over 1194 and 384 citations, respectively \cite{GoogleScholar}, reflecting the extent to which this formulation has permeated the field.  Our preliminary survey indicates that more than 20,000 publications have utilized this equation, often reporting size and strain values based on a formulation that lacks dimensional consistency. The scale of this propagation underscores how a seemingly minor algebraic oversight can spread widely through the scientific literature and influence entire research domains.

Although the incorrect form has been overwhelmingly dominant, a few publications have employed the dimensionally consistent form of the size–strain relation in earlier works; however, its adoption has been limited, and the majority of publications have continued to use the dimensionally inconsistent form. Tatarchuk \textit{et al.}\ \cite{Tatarchuk2017} used a physically valid formulation in their analysis of Zn-doped CoFe$_{2}$O$_{4}$ nanoparticles, although no reference was provided for its derivation, making it unclear how the authors arrived at the dimensionally consistent expression. These isolated cases do not diminish the broader trend, in which the incorrect form has persisted as the dominant version of the SSP equation for more than a decade.

The commonly used version of the SSP equation is written as:

\begin{equation}
	\left(d_{hkl}\beta_{hkl}\cos\theta\right)^2 = \frac{K}{D}\left(d_{hkl}^2\beta_{hkl}\cos\theta\right) + \left(\frac{\sigma}{2}\right)^2,
	\label{eq:zakssp}
\end{equation}
where $d_{hkl}$ is the interplanar spacing (dimension of length, $[L]$), $\beta_{hkl}$ is the corrected line broadening after instrumental subtraction (in radians), $K$ is a dimensionless shape factor (typically $\approx 0.9$, geometry-dependent), $D$ represents the mean size of coherently diffracting domains (typically expressed in nanometres), and $\sigma$ denotes the dimensionless microstrain parameter. In practice, the relation is treated as linear, $y = mx + c$, with
\[
y=\bigl(d_{hkl}\,\beta_{hkl}\cos\theta\bigr)^2,\qquad
x=\bigl(d_{hkl}^{\,2}\,\beta_{hkl}\cos\theta\bigr),
\]
so that the slope $m$ and intercept $c$ are used to estimate $D$ and $\sigma$, respectively.

A straightforward dimensional check exposes an inconsistency. Writing length as $[L]$, the left-hand side of Eq.~(\ref{eq:zakssp}), $\bigl(d_{hkl}\beta_{hkl}\cos\theta\bigr)^2$, has dimension $[L]^2$. The first term on the right-hand side, $\frac{K}{D}\,\bigl(d_{hkl}^{\,2}\beta_{hkl}\cos\theta\bigr)$, has dimension $[L]^1$ (since $K$ is dimensionless, $D$ has the dimension of length $[L]$, and $d_{hkl}^{\,2}$ has the dimension of $[L]^2$), whereas the second term, $(\sigma/2)^2$, is dimensionless, $[L]^0$. 

Thus the equation mixes incompatible dimensions:
\[
[L]^2 = [L]^1 + [L]^0,
\]
violating the principle of dimensional homogeneity.

The implications are immediate:
\begin{enumerate}
	\item Since both plotted variables, $x$ and $y$, carry units of ($[L]^2$), the derived parameter ($D =K/m$) becomes dimensionless ($[L]^0$), rather than having the units of length ($[L]$).
	\item The strain from the intercept, $\sigma = 2\sqrt{c}$, acquires units of $[L]$ rather than being dimensionless ($[L]^0$).
\end{enumerate}
Moreover, the results depend on the arbitrary choice of length units (e.g., \AA{} versus nm), such that merely changing units alters the computed size and strain—a nonphysical outcome. Therefore, any microstructural parameters derived from Eq.~(\ref{eq:zakssp}) are physically meaningless. 

\begin{figure}
	\centering
	\includegraphics[width=0.9\linewidth]{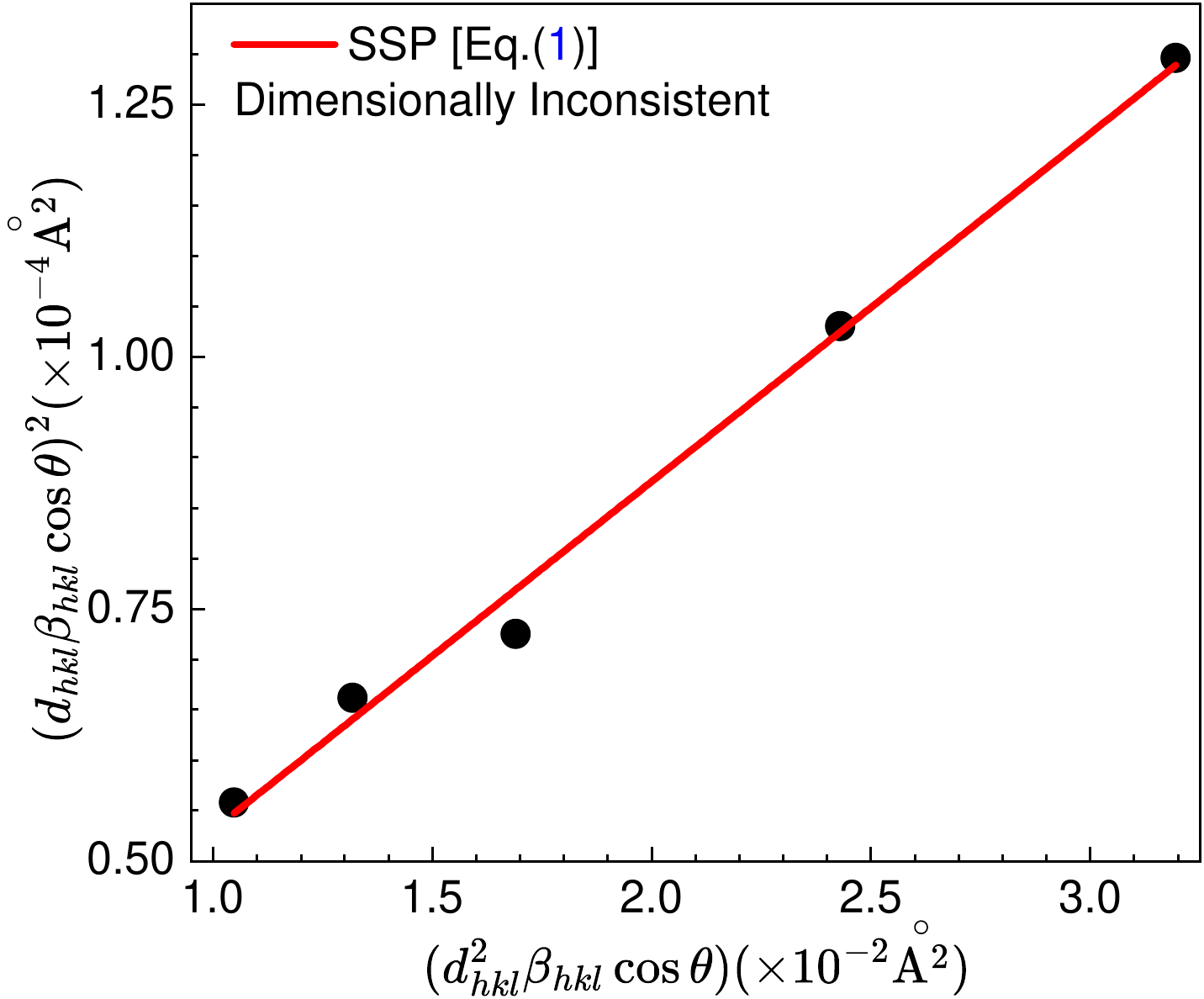}
	\caption{Representative Size--Strain Plot (SSP) constructed using the dimensionally inconsistent formulation [Eq.~(\ref{eq:zakssp})]. The resulting intercept, slope, and the corresponding size and strain parameters are summarised in Table~\ref{tab:comparison_results}.}
	\label{fig:ssp_wrong}
\end{figure}

We fitted Eq.~(\ref{eq:zakssp}) to a representative dataset, as shown in Figure~\ref{fig:ssp_wrong}. The plotted points exhibit an apparently well-defined linear trend, and the fitted parameters fall within numerically reasonable ranges (Table~\ref{tab:comparison_results}). The estimated domain size and microstrain values appear plausible at first glance, mainly because they lie within the expected orders of magnitude for typical nanocrystalline materials. This partly explains why the incorrect relation remained unnoticed for so long: despite its dimensional inconsistency, Eq.~(\ref{eq:zakssp}) frequently yields smooth regressions and plausible values in the numerically expected order of magnitude. 
However, this apparent validity is deceptive. Because Eq.~(\ref{eq:zakssp}) combines terms with incompatible physical dimensions, the yielded size and strain parameters cannot be interpreted as true material properties. The dimensional mismatch fundamentally compromises the physical meaning of the results, even in cases where the statistical fit and estimated parameters appear satisfactory.
Due to this deceptive, consistent behaviour, widespread use of this dimensionally inconsistent formulation has propagated erroneous results across the literature, underscoring the need for a rigorous re-derivation of the SSP method with correct dimensional scaling.

\section{Tracing the Origin of the Misformulated SSP Equation}
Given the extensive use of the erroneous Size–Strain Plot (SSP) equation, identifying its origin is essential to correct the scientific record and restore methodological rigor. A detailed literature survey reveals that nearly all studies employing the dimensionally inconsistent SSP formulation ultimately trace their citation lineage to the work of Tagliente \emph{et al.} \cite{Tagliente2008}. In that study, Tagliente \textit{et al.}~\cite{Tagliente2008} cited the work of Langford~\cite{Langford1992}, who expressed the line-profile analysis (LPA) relations in \textit{reciprocal units} rather than angular units to achieve mathematical clarity and maintain dimensional consistency. As Langford~\cite[pp.~114--115]{Langford1992} noted, expressing breadths in terms of $\beta^{*} = \beta \cos \theta / \lambda$ (see Eq.~(16) in Ref.~\cite{Langford1992}) and $d^{*} = 2 \sin \theta / \lambda$ simplifies the analysis and ensures that all quantities possess consistent dimensions of inverse length. This reciprocal-unit formulation forms the basis for the Halder--Wagner(HW) relation later presented as Eq.~(22) in Langford~\cite[p.~118]{Langford1992}.

Assuming that the Lorentzian component of peak broadening arises solely from finite size,
\[
\beta_C^* = \frac{1}{\epsilon},
\]
and that the Gaussian component originates from microstrain,
\[
\beta_G^* = \frac{\sigma\,d^*}{2},
\]
where $d^*$ is the reciprocal lattice spacing, the Halder–Wagner relation can be expressed as:
\begin{equation}
	\left(\frac{\beta^*}{d^*}\right)^2 = \frac{1}{\epsilon}\left(\frac{\beta^*}{{d^*}^2}\right) + \left(\frac{\sigma}{2}\right)^2\!.
	\label{eq:HalderW}
\end{equation}
It is important to note that $\beta^* = \beta_{hkl}(2\theta)\cos\theta/\lambda$ and $d^* = 2\sin\theta/\lambda = 1/d_{hkl}$ are defined in reciprocal units (inverse length) \cite{Langford1992}. Both $\beta^*$ and $d^*$ correctly possess dimensions of inverse length ($[L]^{-1}$), preserving dimensional homogeneity of Eq. (\ref{eq:HalderW}).Here, $\epsilon$ represents the \emph{apparent size}, which is related to the \emph{true crystallite size} $D$ through a shape factor \cite{Wilson1962pps,Langford1978Scherrer}:
\[
\frac{1}{\epsilon} = \frac{K}{D}.
\]
The true crystallite size is defined as the mean cube root of the crystallite volume. The shape factor $K$ is a dimensionless constant typically close to unity, commonly known as the Scherrer constant. Its value depends on several factors, including the definition of peak breadth, crystallite shape, and crystallite-size distribution \cite{Langford1978Scherrer}. For half-width calculations, $K = 0.9$ can be used as an approximate value. For detailed information on Scherrer constants and their derivation, readers are referred to the work of Langford \emph{et al.} \cite{Langford1978Scherrer}. 

A critical departure from this correct formulation first appeared when Tagliente \emph{et al.}~\cite[see page 1058 Eq.~(3)]{Tagliente2008} reproduced the same relation but redefined the breadth parameter as $\beta^{*} = \beta_{hkl}(2\theta)\cos\theta$, inadvertently omitting the factor $1/\lambda$. This omission altered the physical dimension of $\beta^*$ from inverse length to dimensionless, thereby violating the principle of dimensional homogeneity. Whether this error originated as a typographical oversight or a minor algebraic simplification, its consequences were significant: although the algebraic structure remained superficially similar, the resulting equation was no longer physically meaningful.

This dimensional inconsistency was subsequently propagated and reinforced in numerous highly cited studies, including those by Zak \emph{et al.} (2011), Bindu \emph{et al.} (2014), and Prabhu \emph{et al.} (2014), among many others \cite{Zak2011,Bindu2014,Prabhu2014}. These works adopted the incorrect formulation from Tagliente \emph{et al.} without verifying dimensional consistency and further rearranged the expression by substituting $d^* = 1/d_{hkl}$ along with the now-dimensionless $\beta^*$. This led to the widely used but dimensionally invalid equation:
\[
\left(d_{hkl}\beta_{hkl}\cos\theta\right)^2 = \frac{K}{D}\left(d_{hkl}^2\beta_{hkl}\cos\theta\right) + \left(\frac{\sigma}{2}\right)^2.
\]

The apparent algebraic simplicity of this form likely contributed to its rapid and widespread adoption. Repeated citation and uncritical reuse created a self-reinforcing chain of misinterpretation, embedding the incorrect equation deeply into the scientific literature. 

\section{ Dimensionally Consistent SSP Equation }
By this stage, it should be clear that the correct form of the Size–Strain Plot (SSP) equation can be derived directly from the dimensionally consistent Halder–Wagner relation, or equivalently from Langford’s formulation. Substituting the values of the reciprocal breadth, $\beta^* = \beta_{hkl}\cos\theta/\lambda$, and the reciprocal‐lattice vector, $d^* = 1/d_{hkl}$, into Eq.~(\ref{eq:HalderW}) yields the corrected and dimensionally homogeneous expression:
\begin{equation}
	\left( \frac{d_{hkl}\beta_{hkl}\cos\theta}{\lambda} \right)^2 = \frac{K}{D} \left( \frac{d_{hkl}^2\beta_{hkl}\cos\theta}{\lambda} \right) + \left( \frac{\sigma}{2} \right)^2.
	\label{eq:correctssp_eqn}
\end{equation}
Unlike the widely used, dimensionally inconsistent Eq.~(\ref{eq:zakssp}), Eq.~(\ref{eq:correctssp_eqn}) preserves dimensional homogeneity when the angular breadth $\beta_{hkl}$ is expressed in radians. Both sides of the equation now share the same physical dimension ($[L]^0$), ensuring dimensional consistency between the size‐dependent and strain‐dependent terms. Eq.~(\ref{eq:correctssp_eqn}) corresponds directly to Eq. (22) in J. I. Langford (1992) \cite{Langford1992}, demonstrating the dimensional and analytical equivalence of the corrected Size–Strain Plot and the Halder–Wagner relations, see section~\ref{sec:HWSSP}.

\begin{figure}
	\centering
	\includegraphics[width=0.9\linewidth]{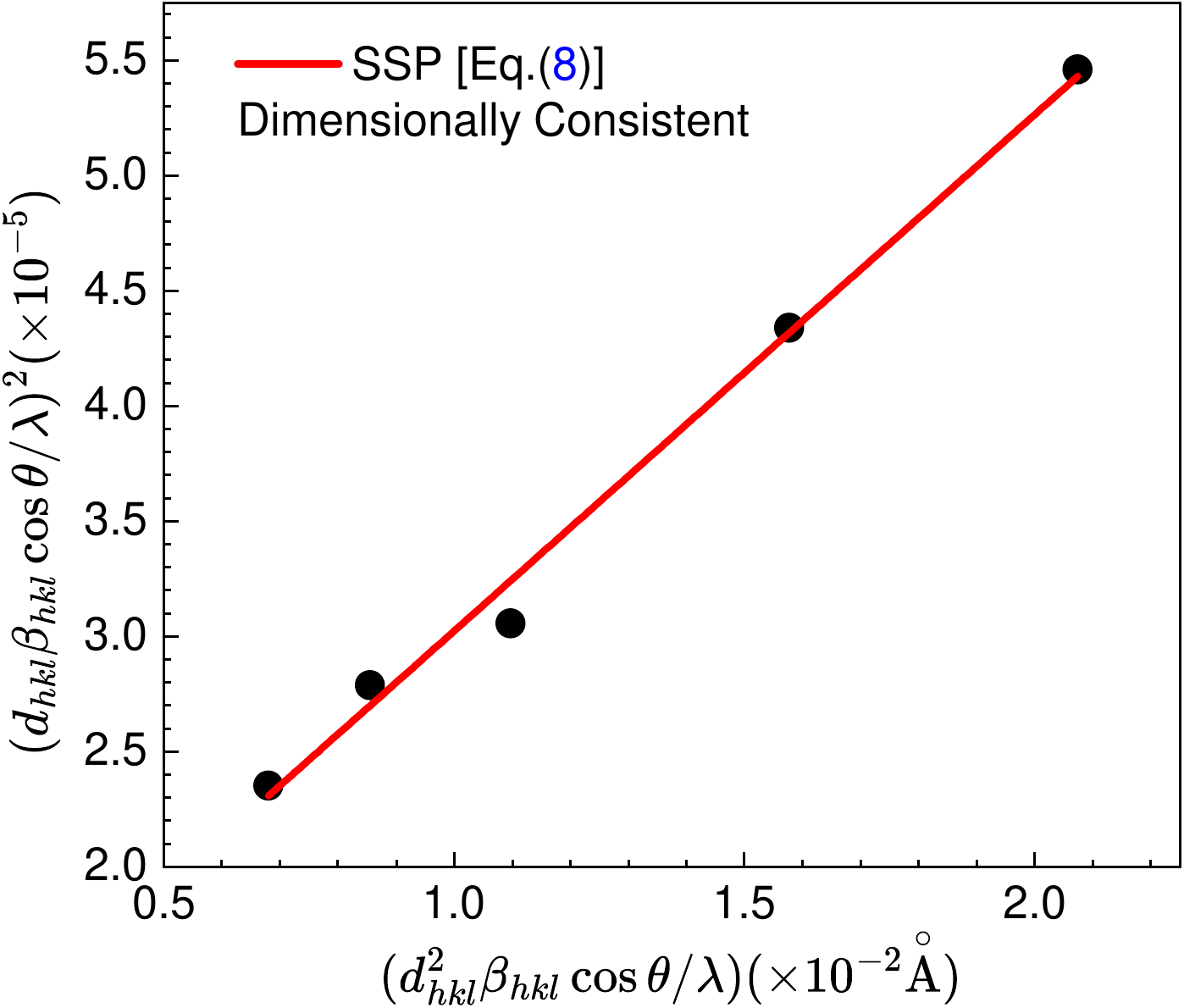}
	\caption{Size--Strain Plot (SSP) constructed using the dimensionally consistent formulation [Eq.~(\ref{eq:correctssp_eqn})]. The fitted intercept and slope, together with the derived domain-size and microstrain parameters, are summarised in Table~\ref{tab:comparison_results}.}
	\label{fig:ssp_eq8}
\end{figure}

We fitted the same experimental data used in Figure~\ref{fig:ssp_wrong} using the dimensionally consistent form of the SSP, as shown in Figure~\ref{fig:ssp_eq8}. In this representation, the linear fit yields a positive slope and intercept, from which physically meaningful domain-size and microstrain values can be obtained. Furthermore, the slope and intercept yield quantities with the dimensions of length and dimensionless, respectively, corresponding to the coherently diffracting domain size and the microstrain. This result is physically consistent and free from the unit-related ambiguities inherent in Eq.~(\ref{eq:zakssp}), thereby demonstrating the correctness of Eq.~(\ref{eq:correctssp_eqn}). The estimated domain-size and microstrain values are summarised in Table~\ref{tab:comparison_results}.

This dimensional consistency is far more than a mathematical formality, it is a requirement for physical validity. It ensures that the SSP relation remains invariant under coordinate transformations and is independent of the units chosen for length or wavelength. As a result, the crystallite size ($D$) and microstrain ($\sigma$) extracted from Eq.~(\ref{eq:correctssp_eqn}) correspond to true material parameters rather than artifacts introduced by dimensional inconsistency.

The corrected formulation, therefore, restores both the mathematical integrity and physical meaning of the SSP method. It eliminates the unit‐dependent errors inherent in the misformulated equation and enables the reliable determination of microstructural parameters from XRD line‐profile data. This reformulated SSP equation re‐establishes the method as a robust and quantitatively accurate tool for microstructural characterization in polycrystalline materials.

\section{Halder–Wagner and Size–Strain Plot Methods}
\label{sec:HWSSP}
The equation presented above [Eq.~(\ref{eq:HalderW})] is widely known as the Halder–Wagner relation. However, it is important to recognize that this expression is fundamentally equivalent to the Size–Strain Plot (SSP) Eq.~(\ref{eq:correctssp_eqn}), as the latter is essentially a simplified graphical representation of the same underlying formulation. Therefore, the apparent distinction between the two methods is historical mainly and representational rather than physical or mathematical.

\begin{figure}
	\centering
	\includegraphics[width=0.9\linewidth]{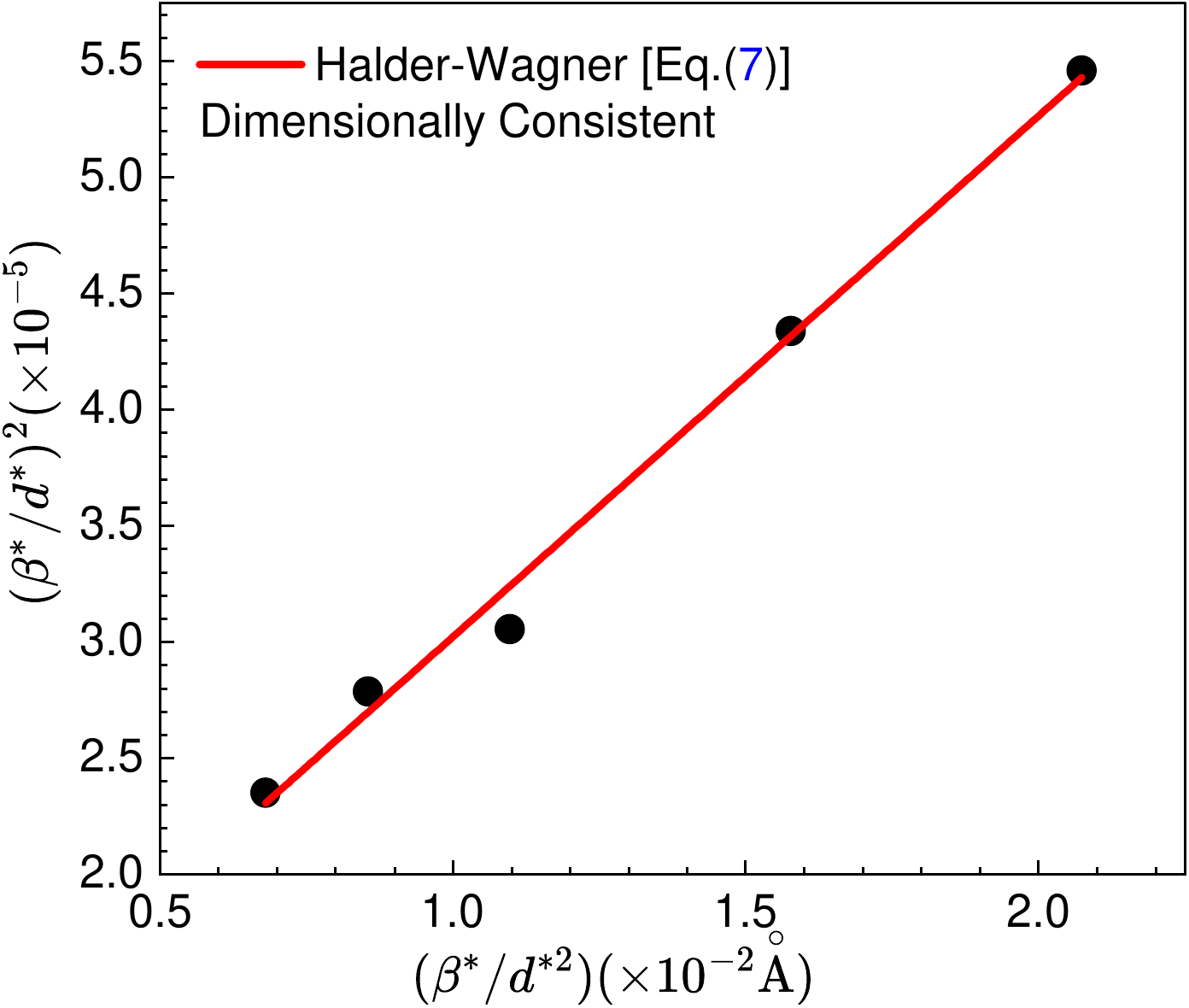}
		\caption{Halder--Wagner plot constructed in reciprocal units according to Eq.~(\ref{eq:HalderW}). The $x$–$y$ values generated by this relation correspond precisely to those of the dimensionally consistent Size--Strain Plot (SSP) in Figure~\ref{fig:ssp_eq8}. Consequently, the fitted trend and all derived parameters are identical, confirming the analytical equivalence of the two formulations. A numerical comparison of the yielded parameters is provided in Table~\ref{tab:comparison_results}.}
	\label{fig:hw_eq7}
\end{figure}

This equivalence highlights a recurring issue in the literature: researchers have frequently presented the same underlying relation under different names and methodological frameworks, creating an impression of novelty without introducing new physical insight \cite{Hassanzadeh-Tabrizi2023, Rabiei2020, Sahu2025, Khalyavka2025}. As a result, both the Halder–Wagner and SSP methods should be understood as two representations of a single analytical approach, differing only in the mathematical form in which they are expressed.
To establish the equivalence between the Halder--Wagner and Size--Strain Plot (SSP) methods, we plotted the same diffraction data using the Halder--Wagner relation [Eq.~(\ref{eq:HalderW})], as shown in Figure~\ref{fig:hw_eq7}. In this representation, the quantities $y = (\beta^{*}/d^{*})^{2}$ and $x = (\beta^{*}/d^{*2})$ were used. Although the HW plot in Figure~\ref{fig:hw_eq7} and the SSP plot in Figure~\ref{fig:ssp_eq8} employ different coordinate variables, the underlying $x$--$y$ data are numerically identical, resulting in the same slope and intercept. Consequently, both methods yield identical values for the coherently diffracting domain size ($D$) and the microstrain ($\sigma$), as summarised in Table~\ref{tab:comparison_results}. This confirms that the HW and SSP approaches are mathematically equivalent representations of the same physical relation, and that the distinction often made between them in the literature is historical rather than analytical.

By clarifying this equivalence, we aim to promote a more coherent and unified use of size–strain analysis within the materials science community. Such clarity is essential to avoid redundancy, prevent conceptual fragmentation, and ensure that future studies build upon a consistent and physically meaningful foundation rather than parallel and superficially distinct formulations of the same result. 

\begin{table*}
	\centering
	\caption{Comparison of the estimated domain-size and microstrain parameters obtained from the three methods: the dimensionally inconsistent SSP equation [Eq.~(\ref{eq:zakssp})], the Halder--Wagner relation [Eq.~(\ref{eq:HalderW})], and the dimensionally consistent SSP formulation [Eq.~(\ref{eq:correctssp_eqn})].}
	\label{tab:comparison_results}
	\begin{tabular}{lcccccl}
		\hline
		\textbf{Method} 
		& \textbf{Equation Reference}
		& \textbf{Intercept ($\times10^{-6}$)} 
		& \textbf{Slope ($\times10^{-3}$)} 
		& \textbf{Domain Size, \(D\) (nm)}
		& \textbf{Microstrain, \(\sigma \) ($\times10^{-3}$)} \\
		\hline
		SSP (Inconsistent) 
		& Eq.~(\ref{eq:zakssp})
		& $1.8\pm0.3$
		& $3.4 \pm0.1$ 
		& $27.3\pm1.3 $ $\ddagger$
		& $8.6\pm0.8$ $\mathsection$ \\
		
		SSP (Consistent) 
		& Eq.~(\ref{eq:correctssp_eqn})
		& $7.8\pm1.4$ 
		& $2.2\pm0.1$ 
		& $40.9\pm2$ 
		& $5.6\pm0.5$ \\
		
		Halder--Wagner 
		& Eq.~(\ref{eq:HalderW}) 
		& $7.8\pm1.4$ 
		& $2.2\pm0.1$ 
		& $40.9\pm2$ 
		& $5.6\pm0.5$ \\
		\hline
	\end{tabular}
	\begin{flushleft}
		\footnotesize{
			$\ddagger$ The derived quantity is dimensionless, which is physically inconsistent with interpreting ($D$) as a crystallite size. \\[2pt]
			$\mathsection$ The parameter carries the dimension of length (nm). 
		}
	\end{flushleft}
\end{table*}

\section{Conclusion}

This study resolves a long-standing dimensional inconsistency in the commonly used Size–Strain Plot (SSP) method. By tracing the origin of the error and examining its subsequent propagation through the literature, we demonstrate that the widely adopted SSP equation lacks dimensional homogeneity and therefore produces size and strain values that are not physically meaningful. Building on the dimensionally consistent Halder–Wagner formulation, we derive a corrected SSP equation that satisfies both physical and mathematical requirements, yielding unit-invariant, reproducible microstructural parameters. 

We further establish the analytical equivalence of the corrected SSP and Halder–Wagner methods, often treated as distinct in the literature, are in fact different algebraic representations of the same underlying relation. Their historical separation reflects differences in representation rather than underlying physics. Recognizing this equivalence is essential to avoid redundancy and to ensure conceptual clarity in future research. Using representative diffraction data, we show that the corrected formulation produces coherent domain sizes and microstrain values that are physically valid, in contrast to the nonphysical outputs obtained from the inconsistent relation. 

Beyond its immediate corrective value, this work underscores a broader lesson about scientific practice: even well‐established methods require continual scrutiny, and even minor errors, if left unchecked, can propagate widely and distort entire analytical traditions. By clarifying the correct formulation and its historical context, we hope this study will help researchers obtain physically meaningful microstructural parameters and foster greater methodological rigor in the use of X‐ray diffraction line‐profile analysis.

\section*{Acknowledgments}
The author gratefully acknowledges Deepika Shanubhogue for asking the critical question that initiated this investigation and for the fruitful discussions that followed.

\bibliographystyle{elsarticle-num} 
\bibliography{JALCOM-SSP} 


\end{document}